\documentclass[pdflatex,sn-basic]{sn-jnl}

\usepackage{graphicx}%
\usepackage{multirow}%
\usepackage{amsmath,amssymb,amsfonts}%
\usepackage{amsthm}%
\usepackage{mathrsfs}%
\usepackage[title]{appendix}%
\usepackage{xcolor}%
\usepackage{textcomp}%
\usepackage{manyfoot}%
\usepackage{booktabs}%
\usepackage{algorithm}%
\usepackage{algorithmicx}%
\usepackage{algpseudocode}%
\usepackage{listings}%
\theoremstyle{thmstyleone}%
\theoremstyle{thmstyletwo}%

\theoremstyle{thmstylethree}%

\begin{document}

\title[Article Title]{Photometric Analysis of TCP J20171288$+$1156589 - WZ~Sge Type Dwarf Nova with Delayed Ordinary Superhumps Emergence}


\author*[1,2]{\fnm{Alexander} \sur{Tarasenkov}}\email{tarasenkov@inasan.ru}

\author[2,3]{\fnm{Sergey} \sur{Shugarov}}

\author[2]{\fnm{Natalia} \sur{Ikonnikova}}

\author[2]{\fnm{Marina} \sur{Burlak}}

\author[4]{\fnm{Sergey} \sur{Nazarov}}

\author[5]{\fnm{Sjoerd} \sur{Dufoer}}

\affil[1]{\orgname{Institute of Astronomy of the Russian Academy of Sciences}, \orgaddress{\street{48 Pyatnitskaya Str.}, \city{Moscow}, \postcode{119234}, \country{Russia}}}

\affil[2]{\orgdiv{Sternberg Astronomical Institute}, \orgname{Lomonosov Moscow State University}, \orgaddress{\street{Universitetsky Pr., 13}, \city{Moscow}, \postcode{119234}, \country{Russia}}}

\affil[3]{\orgname{Astronomical Institute of the Slovak Academy of Sciences}, \orgaddress{\city{Tatransk\'{a} Lomnica}, \postcode{059 60}, \country{The Slovak Republic}}}

\affil[4]{\orgname{Crimean Astrophysical Observatory}, \orgaddress{\city{Nauchny}, \postcode{298409}, \country{Crimea}}} 

\affil[5]{\orgname{Vereniging voor Sterrenkunde}, \orgaddress{\street{Zeeweg 96}, \city{Brugge}, \postcode{8200}, \country{Belgium}}} 


\abstract{We present the results of photometric analysis of WZ~Sge type dwarf nova TCP J20171288$+$1156589. This object exhibited an outburst with a large amplitude of $>7.9$ magnitudes and was observed for over a month. The photometric evolution of the superoutburst was atypical for WZ Sge-type dwarf novae. Periodogram analisys reveals early superhumps with the most probable period of $0.0611\pm0.0001$ days during the initial decline. After a plateau phase of approximately 11 days, ordinary superhumps (likely stage B) emerged with a period of $0.0616\pm0.0001$ days, corresponding to a superhump excess of $\epsilon=0.008$ correspondingly. This delay in the onset of ordinary superhumps is an unusual feature among WZ Sge stars. We evaluated the main parameters of the system: mass ratio $q=M_{RD}/M_{WD}=0.06\pm0.005$, yielding component masses of $M_{WD}\sim1.0\pm0.15M_{\odot}$ for the white dwarf and $M_{RD}=0.06\pm0.01M_{\odot}$ for the donor. The estimated distance to the system is $\sim850$ pc, and the binary separation is $a=0.67\pm0.03R_{\odot}$.}





\maketitle

\section{Introduction}\label{Intro}

Dwarf novae are a class of cataclysmic variable stars, which demonstrate regular outbursts with an amplitude of several stellar magnitudes. They are close binary systems consisting of a white dwarf and a low-mass donor star that fills its Roche lobe. The material of the donor, flowing out through the inner Lagrange point, forms an accretion disk around the white dwarf~\citep{1,Osaki2005,4}. Depending on the morphology of the light curves during outbursts and the regularity of the outbursts, different types of dwarf novae are distinguished. In particular, SU UMa dwarf novae exhibit 2 types of outbursts: normal outbursts with amplitudes of $2-6$ magnitudes and superoutbursts, which usually have amplitudes from several tenths to one magnitude larger (for example, see light curves in~\cite{2024AstL...50..676S, Pavlenko_2025}), than normal outbursts, and last longer (several weeks)~\citep{2}. Superoutbursts are caused by a combination of thermal and tidal instabilities~\citep{1989PASJ...41.1005O}. Among the SU UMa dwarf novae, a subgroup of WZ Sge objects is distinguished, which exhibit a supercycle (the time between two successive superoutbursts) of great duration - years or tens of years, as well as a large superoutburst amplitude - up to 8-10 magnitudes~\citep{Kato15}. 

Some WZ Sge-type stars exhibit repeated outbursts, known as rebrightenings, shortly after the end of their initial outburst. These are short-term increases in the object's brightness, typically lasting 2-3 days and with an amplitude 1-3 magnitudes smaller than the primary outburst. Some stars have been observed to undergo as many as 11 such rebrightenings, but most often, only one rebrightening occurs. A more detailed description of rebrightenings can be found in the articles~\cite{Uemura08, Kato09b, Namekata17, Pavlenko19, SYS21,Lipunov24}. A possible explanation for this phenomenon is given in paper~\cite{MMH15}.

During superoutbursts, so-called superhumps in light curves of SU~UMa-type dwarf novae are observed.  Superhumps are periodic modulations in the light curve with an amplitude of several tenths of magnitude, superimposed on the main trend. The emergence of superhumps is explained by the development of tidally driven eccentric instability in the elliptical accretion disk around the white dwarf~\citep{3}. The period of the superhumps corresponds to the period of beats between the orbital period of the system and the precession period of the accretion disk~\citep{4}. Superhumps are conventionally divided into $"$early$"$, $"$ordinary$"$, and $"$late$"$. Early superhumps are observed only in dwarf novae of the WZ Sge subclass. They have small amplitudes (usually less than 0.1 magnitudes). Their shape is a double wave, and their period is almost equal to the orbital period. They then change to ordinary superhumps, with an amplitude of up to 0.3 magnitudes and a period several percent longer than the orbital period. The optical period usually changes slightly as the superoutburst develops. The evolution of superhumps is discussed in more detail in the following papers:~\cite{1,Osaki2005,4,Kato15,Kato20}.

However, there are stars in which the superhump evolution deviates from the standard. This paper presents the discovery and photometric study of TCP~J20171288$+$1156589 (hereafter referred to as TCP~J2017), a recently discovered WZ Sge-type dwarf nova which showed atypical superhump development.

\section{Discovery of TCP~J2017 and Observational Campaign}\label{Discovery}

TCP~J2017 was discovered by Yuji Nakamura on 2025-07-19 using 0.10-m f/3 refractor equipped with a CMOS camera. The transient had an unfiltered magnitude of $15^{m}.4$, with a limiting magnitude of $16^{m}.5$. Discovery information is available at~\footnote{http://www.cbat.eps.harvard.edu/unconf/followups/J20171288+1156589.html}. The object was also independently detected by the ASAS-SN~\citep{ASASSN1,ASASSN2} survey system on 2025-07-23.35 and designated ASASSN-25de. Alert notice and primary information about TCP~J2017 was acquired using Astro-COLIBRI platform~\citep{5,6}, after which monitoring photometric observations were started.

For photometric monitoring of TCP~J2017 several telescopes were used. Photometric observations at the Crimean Astronomical Station were conducted using the 600-mm RC600 telescope. The telescope is equipped with a back-illuminated 2048x2048 camera with a $13.5-\mu$m pixel size, a field of view of approximately $22^\prime$, and an image scale of $\approx0.7^{\prime\prime}$/pixel. Observations of TCP~J2017 were also carried out with the 500-mm f/8 Astrosib RC500 telescope of the RCO network~\citep{7}. It is equipped with a FLI Kepler 4040 FSI camera with Johnson-Cousins-Bessel $UBVRI$ filters and BACHES echelle spectrograph. The observer can switch between photometric and spectroscopic modes at any time. The photometric mode was used to study TCP~J2017. Also, the 600-mm (f/12.5) Zeiss telescope of Astronomical Institute of the Slovak Academy of Sciences at Star\'{a} Lesn\'{a} equipped with a FLI ML3041 CCD camera was used. The Newton telescope of the Crimean Astrophysical Observatory, with an aperture of 350~mm and a focal length of 1500~mm, equipped with a QHY 600M IMX455 CMOS camera, was also used for observations. 

\begin{table}[h]
\caption{Stellar magnitudes and colours of stars, used for photometry of TCP~J2017}\label{ststars}%
\begin{tabular}{@{}llllllllll@{}}
\toprule
N & U & B & V & R & I & U-B & B-V & V-R & R-I\\
\midrule
1 & 14.962 & 14.766 & 13.998 & 13.549 & 13.194 & 0.196 & 0.768 & 0.449 & 0.355\\
2 & 15.207 & 14.983 & 14.245 & 13.829 & 13.505 & 0.224 & 0.738 & 0.416 & 0.324\\
3 & 14.026 & 14.026 & 13.490 & 13.154 & 12.883 & 0.000 & 0.536 & 0.336 & 0.271\\
4 & 15.565 & 15.502 & 14.825 & 14.416 & 14.079 & 0.063 & 0.677 & 0.409 & 0.337\\
5 & 15.994 & 16.046 & 15.572 & 15.270 & 15.013 & -0.052 & 0.474 & 0.302 & 0.257\\
6 & 15.843 & 15.849 & 15.146 & 14.727 & 14.376 & -0.006 & 0.703 & 0.419 & 0.351\\
7 & 17.199 & 15.895 & 14.558 & 13.856 & 13.317 & 1.304 & 1.337 & 0.702 & 0.539\\
8 & 15.247 & 15.203 & 14.719 & 14.429 & 14.178 & 0.044 & 0.484 & 0.290 & 0.251\\
9 & 16.162 & 15.940 & 15.202 & 14.771 & 14.405 & 0.222 & 0.738 & 0.431 & 0.366\\
10 & 16.338 & 16.317 & 15.624 & 15.211 & 14.859 & 0.021 & 0.693 & 0.413 & 0.352\\
11 & 16.838 & 16.752 & 16.038 & 15.622 & 15.278 & 0.086 & 0.714 & 0.416 & 0.344\\
12 & - & 18.607 & 17.597 & 17.063 & 16.631 & - & 1.010 & 0.534 & 0.432\\
\botrule
\end{tabular}
\end{table}

\begin{table}[h]
\caption{Log of TCP~J2017 observations. The designations of the telescopes are as follows: Z-600 - 600~mm Zeiss telescope (Star\'{a} Lesn\'{a}), CRAO - 350~mm Newton (Crimea), RC600 - 600~mm Ritchey-Chr\'{e}tien (Crimea), K500 - 500~mm Astrosib RC500 (RCO-Kislovodsk)}\label{obslog}%
\begin{tabular}{@{}llllllllllll@{}}
\toprule
JD-2400000 & U & NU & B &  NB & V & NV & R & NR & I & NI & Telescope\\
\midrule
60876.56 & 15.097 & 2 & 15.815 & 2 & 15.752 & 3 & 15.728 & 3 & 15.690 & 3 & Z-600 \\
60877.40 & - & & - & & 15.896\footnotemark[1] & 104 & - & & - & & CR350 \\
60877.40 & 15.265 & 3 & 15.847 & 3 & 15.906 & 303 & 15.815 & 3 & 15.773 & 2 & RC600 \\
60878.37 & 15.401 & 3 & 16.010 & 3 & 16.042 & 342 & 15.948 & 3 & 15.888 & 3 & RC600 \\
60878.42 & - & & - & & 16.043\footnotemark[1] & 56 & - & & - & & CR350 \\
60879.38 & 15.599 & 3 & 16.140 & 3 & 16.177 & 237 & 16.074 & 3 & 16.014 & 3 & RC600 \\
60879.41 & - & & - & & 16.184\footnotemark[1] & 96 & - & & - & & CR350 \\
60880.39 & 15.727 & 3 & 16.283 & 3 & 16.290 & 312 & 16.195 & 3 & 16.123 & 3 & RC600 \\
60882.40 & - & & - & & 16.490\footnotemark[1] & 100 & - & & - & & CR350 \\
60882.46 & 15.993 & 3 & 16.486 & 3 & 16.503 & 181 & 16.372 & 3 & 16.308 & 3 & RC600 \\
60883.35 & - & & - & & 16.598 & 105 & - & & - & & K500 \\
60883.38 & - & & - & & 16.602\footnotemark[1] & 79 & - & & - & & CR350 \\
60883.40 & 16.165 & 3 & 16.636 & 3 & 16.608 & 200 & 16.488 & 3 & 16.411 & 3 & RC600 \\
60884.33 & - & & - & & 16.686 & 135 & - & & - & & K500 \\
60884.39 & 16.248 & 3 & 16.723 & 3 & 16.695 & 166 & 16.553 & 3 & 16.468 & 3 & RC600 \\
60885.34 & - & & - & & 16.795 & 135 & - & & - & & K500 \\
60885.40 & 16.343 & 3 & 16.811 & 3 & 16.799 & 161 & 16.667 & 3 & 16.581 & 3 & RC600 \\
60886.37 & - & & - & & 16.915 & 105 & - & & - & & K500 \\
60887.38 & - & & - & & 16.977 & 105 & - & & - & & K500 \\
60888.37 & - & & - & & 17.087 & 120 & - & & - & & K500 \\
60888.39 & 16.679 & 3 & 17.150 & 3 & 17.092 & 140 & 16.951 & 3 & 16.828 & 3 & RC600 \\
60888.47 & 16.660 & 2 & 17.096 & 3 & 17.094 & 3 & 16.889 & 3 & 16.967 & 3 & Z-600 \\
60892.55 & - & & - & & 17.104 & 2 & 17.030 & 3 & - & & Z-600 \\
60893.38 & 16.701 & 3 & 17.167 & 3 & 17.138 & 3 & 16.993 & 100 & 16.922 & 3 & RC600 \\
60893.41 & 16.730 & 7 & 17.231 & 9 & 17.120 & 9 & 17.053 & 9 & 16.894 & 9 & Z-600 \\
60894.36 & 16.791 & 3 & 17.266 & 3 & 17.204 & 3 & 17.108 & 92 & 17.047 & 3 & RC600 \\
60894.39 & - & & 17.165 & 6 & 17.257 & 6 & 17.166 & 6 & 16.921 & 6 & Z-600 \\
60896.32 & - & & - & & 17.290 & 113 & - & & - & & K500 \\
60896.36 & 16.961 & 2 & 17.459 & 2 & 17.405 & 2 & 17.259 & 2 & 17.233 & 1 & RC600 \\
60898.29 & - & & 18.214 & 5 & 18.388 & 5 & 18.064 & 5 & - & & RC600 \\
60898.39 & - & & - & & 18.098 & 8 & - & & - & & K500 \\
60900.33 & - & & - & & - & & 18.531 & 3 & - & & RC600 \\
60900.34 & - & & - & & 18.370 & 13 & - & & - & & K500 \\
60902.40 & - & & - & & - & & 19 \footnotemark[2] & 3 & - & & Z-600 \\
60907.40 & - & & - & & - & & 20 \footnotemark[2] & 3 & - & & Z-600 \\
\botrule
\end{tabular}
\footnotetext[1]{Unfiltered passband reduced to V.}
\footnotetext[2]{Limiting magnitude.}
\end{table}

Also AAVSO observations were used for our analysis. These observations were obtained with the Fregenal de La Sierra (E-Eye) 40cm f/6.8 corrected Dall-Kirkham telescope in Spain, equipped with camera: QHY268M in unfiltered passband.

\begin{figure*}[h]
\centering
\includegraphics[width=0.48\textwidth]{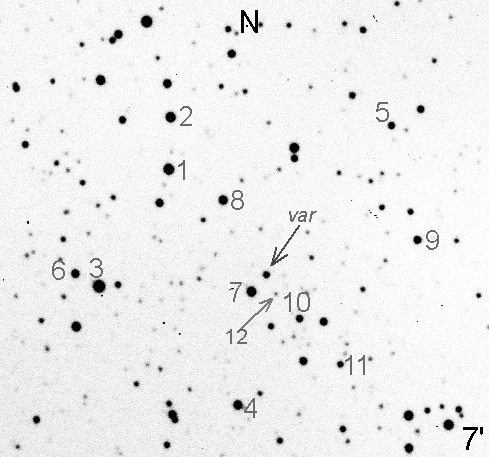}
\includegraphics[width=0.39\textwidth]{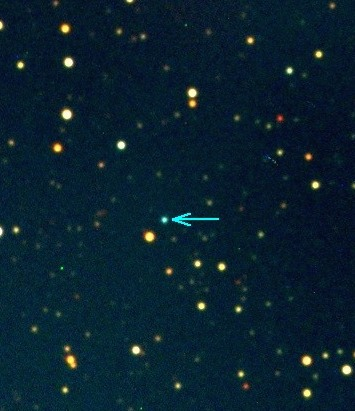}
\caption{Finding charts for TCP~J2017: left panel indicates TCP~J2017 (marked as var) and comparison stars; right panel shows combined colour image, TCP~J2017 is marked with an arrow and has a noticeable blue colour. }\label{map}
\end{figure*}

Photometry of the frames was carried out using the programs MaximDL and AstroimageJ~\citep{8}. For a photometric study of the object, the $UBVRI$ magnitudes of 12 nearby stars were determined using the Z-600 telescope (Slovakia) (see Figure~\ref{map}). Magnitudes and colours of these stars are presented in Table~\ref{ststars}. The magnitudes were obtained by referencing the stars to a photometric standard near PU Vul~\citep{HM06} over four nights. Note that the magnitudes of the stars around PU Vul were obtained by referencing the stars measured by Landolt~\citep{Landolt92}. The same set of comparison stars was used for photometry of TCP~J2017 on frames obtained by all telescopes. The data from all telescopes were converted to a single system, as close as possible to the standard Johnson-Cousins system, and corrected for systematic errors associated with differences in the instrumental systems of different instruments. However, we will designate this system as $UBVRI$ further in the text. Log of photometric observations is shown in the Table~\ref{obslog}. 

\section{Historical Light Curve}\label{HistLC}

TCP~J2017 in the quiescent state is absent from the Gaia DR3~\citep{9} catalogue, the ZTF~\citep{10} observation archive (accessed via SNAD web tool~\citep{11}), and the Pan-STARRS~\citep{12} catalogue. Given that Pan-STARRS has a limiting magnitude in the $g$ filter of $23^{m}.3$, and TCP~J2017 at the peak of the outburst reached $15^{m}.4$, we assume the lower limit of the outburst amplitude to be about $7^{m}.9$, which is a high value even for WZ~Sge-type dwarf novae. ASAS-SN~\citep{13,14} shows no evidence of previous outbursts over the past 13 years.

\section{Observed Light Curve Features}\label{ObsLC}

\begin{figure}[h]
\centering
\includegraphics[width=0.9\textwidth]{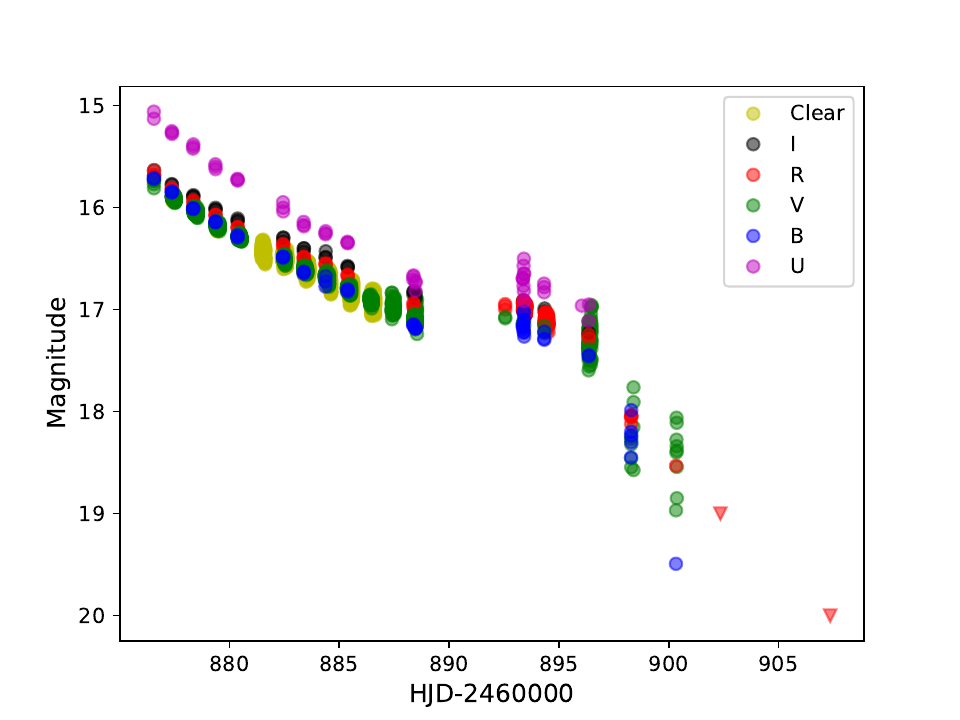}
\caption{Full lightcurve of TCP~J2017. Data points are represented as circles and photometric limits - as triangles.}\label{fulllc}
\end{figure}

\begin{figure}[h]
\centering
\includegraphics[width=0.9\textwidth]{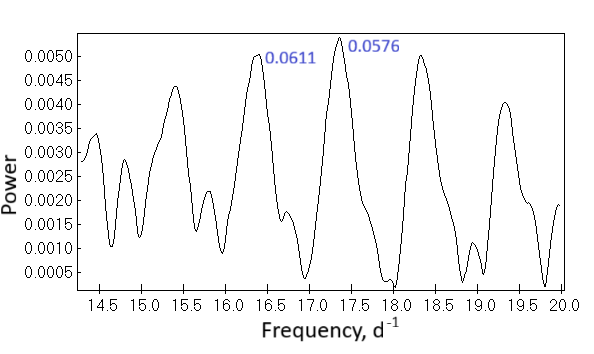}
\caption{Periodogram of the light curve of TCP~J2017 for the first 4 nights. Peaks corresponding to probable periods of early superhumps are marked.}\label{Period}
\end{figure}

\begin{figure*}[h]
\centering
\includegraphics[width=0.45\textwidth]{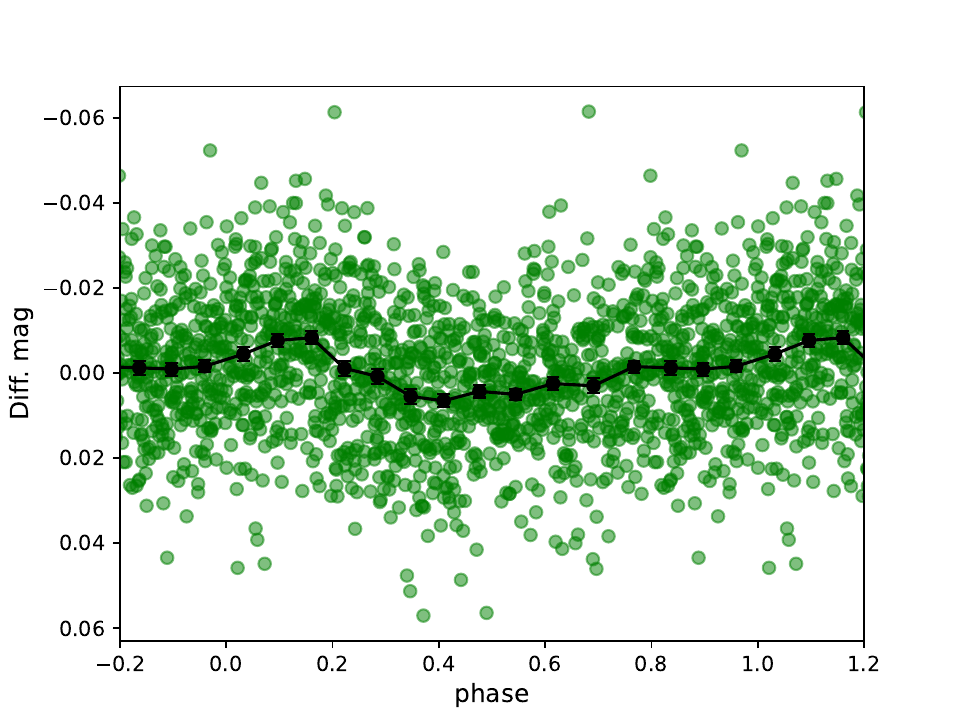}
\includegraphics[width=0.45\textwidth]{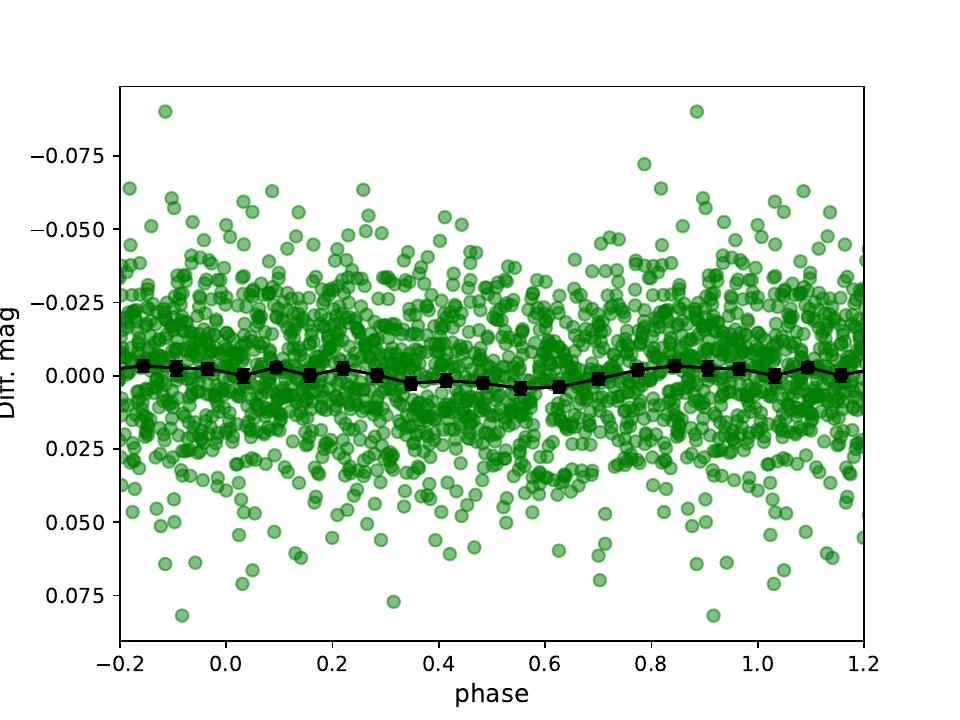}
\includegraphics[width=0.45\textwidth]{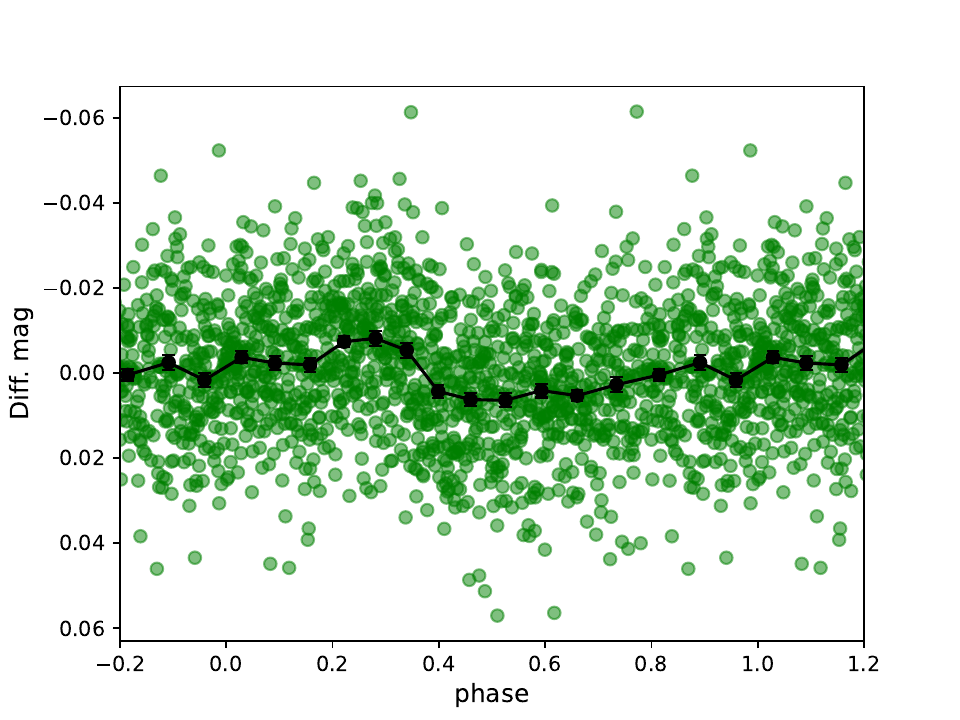}
\includegraphics[width=0.45\textwidth]{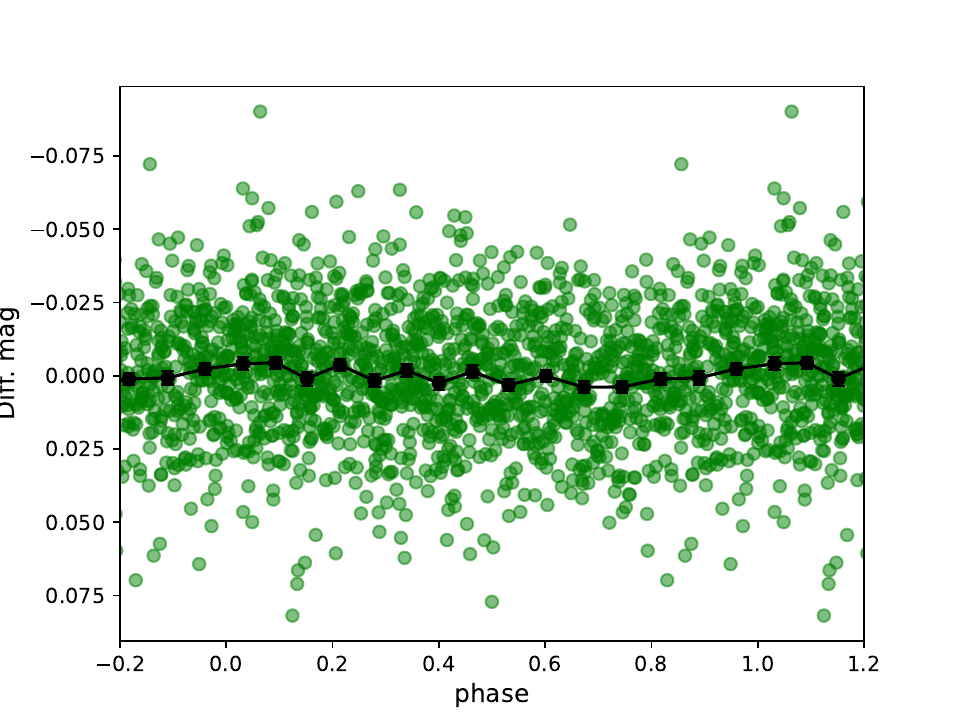}
\caption{Profiles of early superhumps of TCP~J2017, folded with period $0.0576\pm0.0001$ days (upper panels) and $0.0611\pm0.0001$ (lower pannels): first 4 nights (left panels) and other nights of primary decline phase (right panels). Data points are green circles, the average profile is shown by black connected dots.}\label{EarlySH}
\end{figure*}

TCP~J2017 was observed for a month after its discovery, resulting in a multi-colour $UBVRI$ light curve (See Figure~\ref{fulllc}). During about the first two weeks  TCP~J2017 demonstrated gradual fading at a rate of ~ $0.1$ mag/day. Colour indices during this phase of outburst are: $B-V$ = $-0^m.03$, $V-R$ = $0^m.1$, $R-I$ = $0^m.06$, which are characteristic for dwarf nova outbursts. In early stage of outburst TCP~J2017 has significant UV excess ($U-B=-0^m.58$), which decreases during outburst. To search for possible periodicities in the monotonic decay phase, the light curve combined from all telescopes was detrended by subtracting the polynomial trend. The search for periodicities was carried out using the WinEFK program using the Deeming~\citep{15} and Lafler-Kinman~\citep{16} methods. Analysis reveals a low-amplitude periodic modulation with a period of $0.0576\pm0.0001$ or $0.0611\pm0.0001$ days. The periodogram with peaks corresponding to the identified periods is shown in Figure~\ref{Period}. A detailed justification for why we chose these periods for analysis, rather than those corresponding to other high peaks of the periodogram, is given in Section~\ref{Res}. This trend was present during the first 3-4 nights of monitoring and had an amplitude of about $0^m.02$ in $V$ passband. In the following nights of the monotonic decline stage, the trend disappeared. The folded light curves for two intervals (JD 2460876–2460880 and 2460882–2460888) are shown in Figure~\ref{EarlySH} (left and right panels, respectively). For clarity, the averaged profile is overplotted on the observed points. We interpret this sinusoidal variations as early superhumps.

\begin{figure}[h]
\centering
\includegraphics[width=0.9\textwidth]{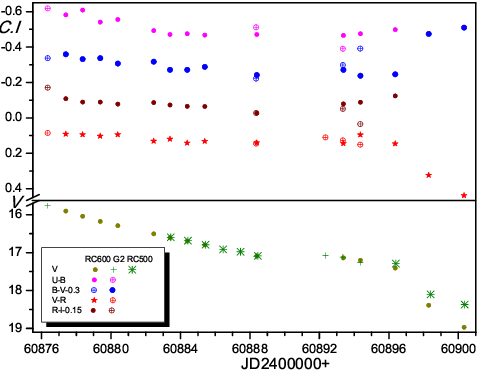}
\caption{Colour index curves for TCP~J2017 (upper panel). $V$ passband light curve is shown in the lower panel.}\label{colorsplot}
\end{figure}

Figure~\ref{colorsplot} shows graphs of changes in colour indices from time. For better visualization on the lower part of Figure~\ref{colorsplot} the light curve in the $V$ passband is shown, where only measurements in the filter, averaged over every night are plotted. One can see a small (not more than 0.2 magnitudes) increase in all colour indices in the initial area of the fall of brightness (plateau). This is caused by gradual cooling, a decrease in the average temperature of the emitting areas of the accretion disk, a hot spot on it, and an accretion stream. These features make the main contribution to the total radiation. However, at the end of the plateau, the $B-V$ colour index began to decrease again. A similar effect was observed for V1006 Cyg~\citep{Pavl24}. We explain this change by the shrinking of the accretion disk, which exposed the hot white dwarf and allowed it to contribute to the system’s total flux. Also, this effect can be explained by the re-appearance of accretion disk emission lines, which are suppressed at the initial stages of the outburst, in the outburst final stages. The Balmer lines contribute much more to the flux in the $B$ passband compared to the $V$ passband, leading to the observed "blueing" of the $B-V$ colour index. Differences in the spectra of dwarf novae in the outburst and quiescent state are shown, for example, in~\cite{2020PASJ...72...76H, 2018AstBu..73...84V}. Similar changes in colour indices occur in many SU UMa and WZ Sge-type stars (see~\cite{Krush24,Marina,SYS15,Chochol15}). We will describe the nature of these changes in more detail in Section~\ref{Colours}.

\begin{figure*}[h]
\centering
\includegraphics[width=0.45\textwidth]{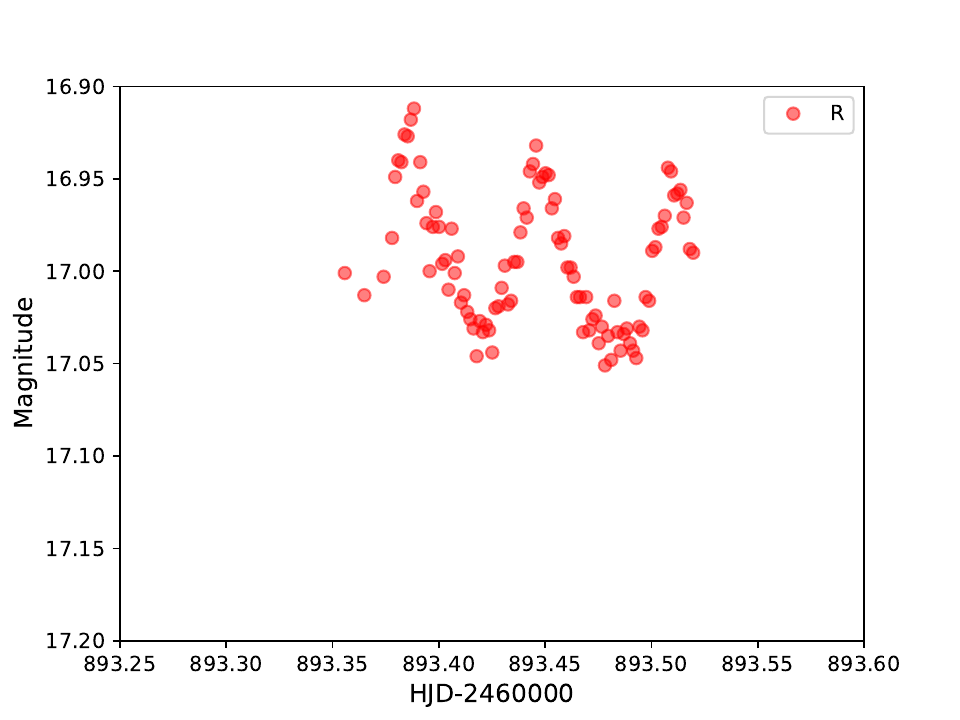}
\includegraphics[width=0.45\textwidth]{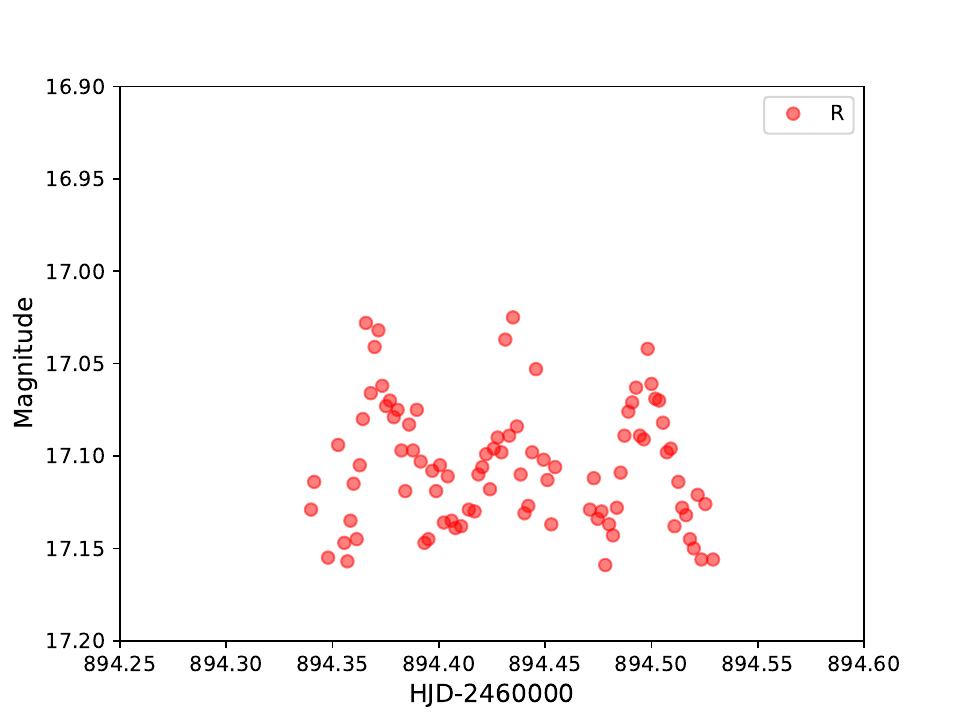}
\caption{Ordinary superhumps of TCP~J2017: 2025 August 5th (left panel) and 2025 August 6th (right panel).}\label{SHlc}
\end{figure*}

\begin{figure}[h]
\centering
\includegraphics[width=0.9\textwidth]{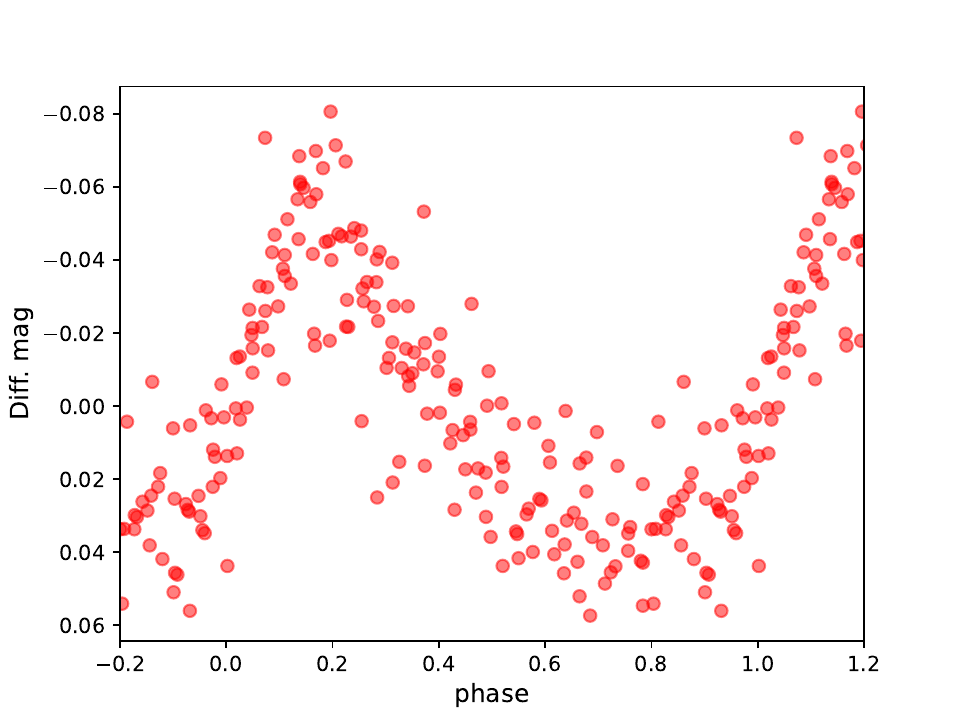}
\caption{Ordinary superhumps of TCP~J2017 folded with period $0.0616\pm0.0001$ days.}\label{SHfold}
\end{figure}

After JD 2460888, the light curve reached a roughly constant brightness level ($17-17.2^m$) and ordinary superhumps began to develop. The most pronounced ordinance superhumps were observed on JD 2460893 and 2460894 (the 17th and 18th days after the start of observations). They had a characteristic sawtooth shape and an amplitude of $0^m.10-0^m.15$ in the $R$ passband. The light curves for these two nights are shown in the Figure~\ref{SHlc}. Based on observations during these 2 nights, we determined the superhump period to be $0.0616\pm0.0001$ days. The superhumps profile folded with this period is shown in the Figure~\ref{SHfold}.

After this stage, at JDs later than 2460896, the brightness of TCP~J2017 began to decline rapidly at a rate of approximately 1 mag/day, and the object quickly faded below the detection limit of the telescopes used for the observations. No rebrightenings were detected after the main outburst. However, we do not exclude the possibility that TCP~J2017 demonstrated the short-term decrease in brightness during plateau phase (so called "dip"), which is observed in some SU UMa-type dwarf novae~\citep{2016PASJ...68...55K,2023arXiv230407695K}. This dip could be missed due to unfavourable weather conditions for observations.

A similar atypical development of superhumps was also observed in the star PNV J1842+4837 (V529 Dra)~\citep{V529dra}. This star also exhibited early superhumps in the first days after the outburst, after which the consistent periodic brightness fluctuations disappeared. Ordinary superhumps appeared only at the end of the outburst, during the rebrighting process. However, it can be assumed that this star did not undergo a rebrighting process, but rather a brief brightness drop of 5 magnitudes, after which the star's brightness returned to almost the same level and the monotonic brightness decline continued. No such brightness decline was observed in the star we studied. However, we do not rule out that it was missed due to bad weather conditions and its short duration.

\section{Colour Variations}\label{Colours}

Analysis of the positions and tracks of cataclysmic variable stars in two-colour diagrams can provide important information about certain physical processes occurring in the system. The tracks were calculated theoretically in papers~\cite{Mayo80,Ech83,Cann87} and others. The authors accounted for the variable emission from the accretion disk using the law found by~\citep{SS73}, as well as taking into account various emission sources in the system at different stages of the outburst and the influence of emission lines to calculate the integrated flux in the $UBVRI$ passbands. Theoretical tracks and loops were plotted on various two-colour diagrams, which showed good agreement with the observational results~\citep{Bruch84,Ech1984}.

One of the goals of our work was to plot the tracks of TCP~J2017 on two-colour diagrams and interpret them.

\begin{figure*}[h]
\centering
\includegraphics[width=0.3\textwidth]{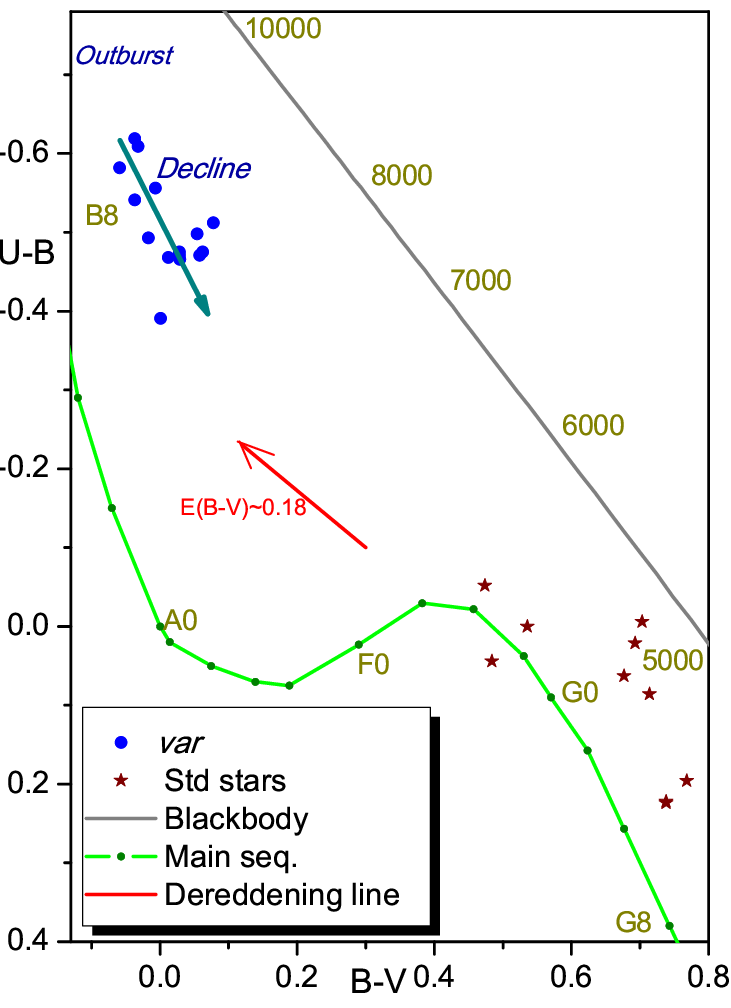}
\includegraphics[width=0.3\textwidth]{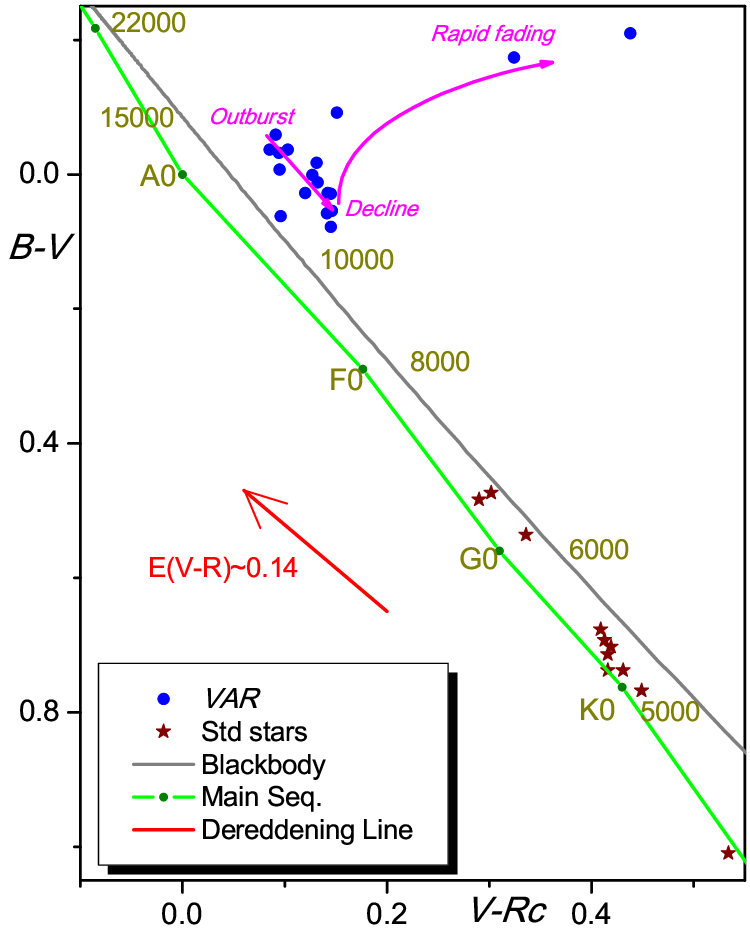}
\includegraphics[width=0.3\textwidth]{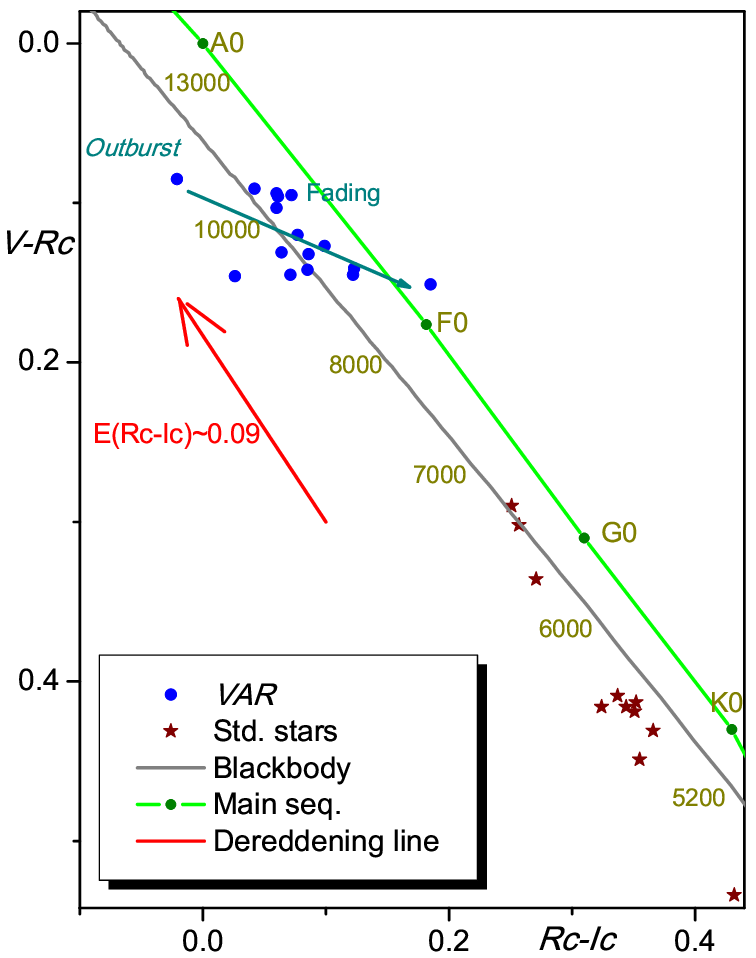}
\caption{Two colour diagrams of TCP~J2017: (\textbf{a}) $U-B$-$B-V$ (\textbf{b}) $B-V$-$V-R$ (\textbf{c}) $V-R$-$R-I$.}\label{ColDiag}
\end{figure*}

In Section~\ref{ObsLC}, we already showed that the object's colour indices change slightly during outbursts and brightness declines. To study these changes in more detail, we plotted the object's positions and tracked its movement on two-colour diagrams (see Figure~\ref{ColDiag}).

The $U-B$/$B-V$ diagram plots the positions of the field stars from Table~\ref{ststars} (see left panel of Figure~\ref{ColDiag}). The interstellar reddening value, $E(B-V)$, calculated using various extinction calculators~\footnote{http://argonaut.skymaps.info/query}~\footnote{https://astro.acri-st.fr/gaia\_dev/\#extinction}~\footnote{https://ned.ipac.caltech.edu/forms/calculator.html}, ranges from $0.11$ to $0.23$ for distances between $0.6$ and $1$ kpc. Figure~\ref{ColDiag} (left panel) shows that the average reddening value is $E(B-V) = 0.18$. 

Most of the stars we measured are most likely main-sequence (MS) stars of spectral class F-G. The distance to these stars can be estimated as no more than $1$ kpc. The fact that nearer stars exhibit less reddening is consistent with the location of neighbouring stars on the two-colour diagram. However, if giants are present among these stars, their distance and reddening would be greater. According to~\cite{Kato15} (see Chapter 6), the absolute magnitude of dwarf novae during an outburst is approximately $+5^m$, which is comparable to that of F-G main-sequence stars. Therefore, considering the range of probable reddening values from the mentioned calculators and our own approximate estimate, we adopted a mean value of $E(B-V) = 0.18$.


The position of TCP~J2017 and the track from the outburst to the onset of decay are shown by blue symbols and a dark green arrow. If we shift this position by the interstellar extinction value, the average colour temperature of the source will vary from $\sim15,000K$ at the beginning of the outburst to $\sim10,000K$ by the end of the plateau. We did not conduct later observations, during the dramatic fading and final decline in the $U$ passband.

Middle and right panels of Figure~\ref{ColDiag} similarly show the positions of the comparison stars and TCP~J2017. We conducted multicolour observations not only during the outburst and plateau, but also during the rapid and final brightness decline. During this decline, the $B-V$ colour index of TCP~J2017 decreased by approximately 0.2 mag compared to the value at the beginning of the outburst. The $V-R$ and $R-I$ indices continued to increase. 

After the outburst, the accretion disk's size decreases, its temperature and luminosity drop, and its colour indices increase. We assume that the white dwarf was obscured by the accretion disk during the outburst, preventing its radiation from reaching the observer. However, by the end of the outburst, we suppose that the disk's size has decreased, and the hot white dwarf has become visible. In this case, fluxes in the blue and UV spectral regions should increase due to the additional light from the white dwarf and also due to the manifestation of emission lines that were suppressed during the outburst. We indeed observe similar behaviour in the blue spectral region ($B$ passband) (track to the right in the middle panel of Figure~\ref{ColDiag}, but we have no observations in the $U$ passband. In the $VRI$ photometric passbands, the contribution from the hot white dwarf is small, with its maximum contribution occurring in the UV. Therefore, the cooling disk continues to increase the $"$redder$"$ $V-R$ and $R-I$ colour indices, as seen in the right panel of Figure~\ref{ColDiag}.

The average colour temperature we found during the outburst, 10,000-15,000K, is higher than the average for similar stars. Most SU and WZ dwarf novae have average temperatures between $8,000$ and $12,000$K. Some of these stars are ASASSN-19oc~\citep{Krush24}, AY Lac~\citep{Marina}, MO Psc~\citep{SYS15}, OV Boo~\citep{OVBoo}, RZ LMi~\citep{RZ18}, V1838 Aql~\citep{V1838}, etc. However, there are also hotter stars, with temperatures from $12,000$ to $20,000$K, these are CV Tri~\citep{CVTri}, V627 Peg~\citep{V627}, PR Her~\citep{SYS15} and other objects.

We do not rule out the presence of a small systematic error in the determination of stellar magnitudes, although the calibration was carried out over several nights. This error may primarily affect the $I$ passband. The long-wavelength cutoff of this band is determined by the sensitivity roll-off of the specific CCD chip, which varies among different CCD models and may potentially change over time, rather than being strictly constant. This is evident in the right panel of Figure~\ref{ColDiag}.

It is impossible to correct the positions of the stars for interstellar reddening in such a way that they would lie precisely on the main sequence. However, this error does not exceed $0^m.05$ and does not alter our main conclusions or the estimates of the system parameters.

\section{Main Results}\label{Res}

For TCP~J2017, we found two possible periods of the early superhumps - $0.0576\pm0.0001$ and $0.0611\pm0.0001$ days. This period is very close to the orbital period in similar systems~\citep{4,Kato15}. At the end of the outburst, we detected the period of ordinary superhumps ($0.0616\pm0.0001$ days), probably in the $"$B$"$ stage.

Since TCP~J2017 is absent in the Gaia catalogues, the distance cannot be calculated using parallax, but it can be estimated using the standard formula
\begin{equation}
\rm d = 10^{0.2\cdot(V-M_v+5-A_v)},
\end{equation}
where $A_v$ is the extinction in the $V$ passband, $V$ is the peak magnitude during the outburst in the $V$ passband, $M_v$ is the absolute magnitude of the object during the outburst. We put the value $V\approx15.5$ based on the magnitude of the object at the time of detection and observations at early outburst stage (see Table~\ref{obslog}). According to~\cite{Kato15}, the absolute magnitude at the maximum of a superoutburst of SU UMa-type stars is 
\begin{equation}
M_v (max) = 5.64–0.259 \times P_{orb}(hr).
\end{equation}
For $P_{orb}$ of most stars of the WZ Sge-type systems and also TCP~J2017 this formula gives $M_v (max) = 5.3$. So, $V – M_v = 10.2$. The value of interstellar extinction $A_v = 3.1E(B - V)\sim0.55$ mag. As a result, we get a distance to the system TCP~J2017 $d\approx850$ pc.

With known both the orbital period ($P_{orb}$) and the period of the ordinary superhumps ($P_{ord}$), it becomes possible to calculate the superhump excess value ($\epsilon$): 
\begin{equation}
\epsilon = (P_{ord}/P_{orb})-1.
\end{equation}
In the case of TCP~J2017 $\epsilon = (0.0616/0.0576) - 1 = 0.07$ or $\epsilon = (0.0616/0.0611) - 1 = 0.008$ for two possible early superhumps periods. This values, especially the second one, are typical for SU UMa and WZ Sge-type stars (see right panel of Figure 19 in~\cite{4} for the distribution of $\epsilon$ for SU UMa stars). The mass ratio of stellar components $q$ ($q = M_{RD}/M_{WD}$, where $M_{RD}$ is the mass of the donor star, and $M_{WD}$ is the mass of the white dwarf) can be estimated from the value of $\epsilon$. According to~\cite{Kato22}, we get $q=0.29\pm0.03$ or $q=0.06\pm0.005$ respectively.

Masses of red dwarf in TCP~J2017 system can be approximately estimated from semi-empirical sequence as $0.11 \pm 0.03 M_{\odot}$ or $0.06 \pm 0.01 M_{\odot}$ respectively (see Tables 2, 6 and Figure 9 in~\cite{Knigge11}). Since we found the value $q = M_{RD}/M_{WD}$ for TCP~J2017, the mass of the white dwarf in this system will be equal to $M_{WD}\sim0.4\pm 0.15 M_{\odot} $ or $M_{WD}\sim1.0\pm 0.15 M_{\odot}$ correspondingly.

From the component masses and orbital period, the distance between the components in the system can be obtained using Kepler's third law, modified by Newton. Thus, $a = 0.50 \pm 0.03 R_{\odot}$ in case of the first period and $a = 0.67 \pm 0.03 R_{\odot}$ in case of the second period.

Since WZ Sge-type systems are typically characterized by a mass ratio $q$ of $0.07-0.08$~\citep{Kato15}, that's why we consider the period of early superhumps of $0.0611\pm0.0001$ days, corresponding to the value of $q=0.06\pm0.005$, represents the actual behaviour of the system. The period $0.0576\pm0.0001$ days, which, although giving a slightly higher peak on the periodogram, is 1-day alias to the true one. Its obvious manifestation is due to the fact that all the instruments used are at similar longitudes, so the light curve has regular gaps that occur during daytime. Other alias periods were not considered because they yield values of the system parameters that are in principle impossible for dwarf novae.

\section{Discussion}\label{Discuss}

TCP~J2017 is clearly a WZ Sge dwarf nova, as it exhibited an outburst with an amplitude of at least 7.9 magnitudes, during which early and ordinary superhumps were detected, and archival observations have not revealed previous outbursts for at least the last 13 years. However, its photometric behaviour during outburst is quite unusual. After the presence of early superhumps during the first 3-4 days, a phase of monotonous decline lasting more than 10 days before the emergence of ordinary superhumps sets in. Such a long waiting time between early and ordinary superhumps stages is not typical for outbursts of WZ Sge-type dwarf novae. Such photometric behaviour usually indicates an extremely energetic flare. In particular, the object MASTER OT J030227.28+191754.5, which demonstrates the most delayed occurrence of ordinary superhumps (30-32 days after outburst start) had an outburst amplitude of  10.2 magnitudes~\citep{Tampo24}. The authors suggest that such a large amplitude could be due to a massive,  extremely low-viscosity disk, as well as a possible low inclination of the system. Another similar dwarf nova, GOTO065054+593624 (an ordinary superhumps event occurred 13.8 days after discovery), had a outburst amplitude of 8.5 magnitudes~\citep{GOTO}. Considering the large amplitude of TCP~J2017 ($>7.9$ magnitudes) and low donor mass $M_{RD} = 0.06 \pm 0.01 M_{\odot}$, we assume a similar high-amplitude outburst mechanism of this object. 

However, the TCP~J2017 is distinguished by the absence of dips in the light curve, which often appear in the light curves of dwarf novae of the WZ Sge type. Although in our case the narrow dip could have been missed due to bad weather, its possible absence does not contradict the high amplitude of the outburst. Thus, MASTER OT J030227.28+191754.5 did not demonstrate a dip during the outburst. System TCP J18173469+1803499 (with an amplitude of 8 magnitudes) also did not exhibit a dip, although in this case, ordinary superhumps developed more quickly - on the 8th day after detection~\citep{1817}.

\section{Conclusion}\label{Concl}

We present a comprehensive photometric study of the dwarf nova TCP J20171288+1156589 in outburst. Our analysis allows us to confidently classify this object as belonging to WZ Sge subgroup. The key parameters supporting this classification include a large outburst amplitude of more than $7.9$ magnitudes and the presence of early superhumps with a period of $0.0611 \pm 0.0001$ days in the early stages of the outburst. The period of early superhumps in WZ Sge-type dwarf novae is very close to the orbital period of the system $P_{orb}$. Therefore, we consider them to be approximately equal and use them to estimate other fundamental parameters of the system.

The object exhibited an atypical superhump evolution. The early superhumps, observed during the first $3-4$ nights of the outburst, ceased, and were followed by a plateau in the light curve. Ordinary superhumps, which we associate with stage B superhumps, appeared only after approximately 11 days, on JD~2460893 and JD~2460894, with a period of $P_{ord} = 0.0616 \pm 0.0001$ days. This delay in the onset of ordinary superhumps relative to the early superhump phase is an unusual feature among WZ Sge-type stars.

From the observed periods, we calculated the superhump excess to be $\epsilon = (P_{ord}/P_{orb}) - 1 = 0.008$. This value is typical for SU UMa and WZ Sge-type dwarf novae. This excess corresponds to system components mass ratio of $q = M_{RD}/M_{WD} = 0.06 \pm 0.005$. Using empirical relations, we estimated the component masses to be $M_{RD} = 0.06 \pm 0.01 M_{\odot}$ for donor, which appears to be a brown dwarf, as in some other WZ Sge-type systems~\citep{Kato15,Lipunov24}, and $M_{WD} \sim 1.0 \pm 0.15 M_{\odot}$ for the white dwarf primary.

The estimated distance to the system is $d \approx 850$ pc. Applying Kepler's third law to the component masses and the orbital period, we derived the separation between the components in the binary system to be $a = 0.67 \pm 0.03 R_{\odot}$.

In summary, TCP J2017 is a WZ Sge-type dwarf nova that displays a non-typical superhump evolution, characterized by a significant delay between the early superhump phase and the development of late, ordinary superhumps. This makes it an interesting object for further studies aimed at understanding the accretion disk dynamics in extreme mass-ratio systems.

\backmatter

\bmhead{Data Availability Statement}

All observational data are available upon reasonable request to the corresponding author.

\bmhead{Acknowledgements}

The observations were performed at telescope Astrosib RC500 of shared research facility “Terskol observatory” of Institute of Astronomy of the Russian Academy of Sciences. The observations on RC600 telescopes of SAI MSU were carried out using the equipment purchased under the MSU Program of Development. AstroColibri was used to receive alert notice. The authors are grateful to Dr. V. P. Goranskij for providing the WinEfk program. The study was conducted under the state assignment of Lomonosov Moscow State University. A. Tarasenkov acknowledges the support of the Foundation for the Development of Theoretical Physics and Mathematics BASIS (project 25-2-1-39-1). The authors would like to thank the anonymous reviewer who provided useful and detailed comments which helped improve the paper.

\bmhead{Funding}

Part of this work was supported by the Slovak Research and Development Agency under contract No. APVV-24-0160 and by Vedeck\'{a} grantov\'{a} agent\'{u}ra (VEGA) No 2/0003/25 (S.Y.S.) 

\bmhead{Conflict of interest}

The authors declare no conflict of interest.


\bibliography{biblio}

\end{document}